\documentclass[article,reqno]{amsart}
\usepackage{amsmath,amssymb}
\usepackage{latexsym}
\usepackage{epsfig}
\usepackage{cite}
\usepackage{booktabs}
\usepackage{color}
\usepackage{graphicx}
\usepackage{tikz}
\usepackage{pgfplots}
\usepackage{amsthm}
\usepackage{braket}
\numberwithin{equation}{section}

\newcommand{\beq}{\begin{equation}}
\newcommand{\eeq}{\end{equation}}

\title{De Sitter magnetic black hole dipole with a supersymmetric horizon}
\author{D.Astesiano and S.L. Cacciatori}
\address{Dipartimento di Scienza e Alta Tecnologia, Universit\`a dell'Insubria, Via Valleggio 11, 22100 Como, Italy, and INFN, Via Celoria 16, 20133 Milano, Italy}

\begin{document}

\begin{abstract}
We find a new non BPS solution in $N=2$ $D=4$ gauged supergravity coupled to $U(1)$ gauge fields and matter. It consists in a closed universe with two extremal black holes of equal size, surrounding two singularities. They  
have opposite magnetic charges (and no electric charges), but stay in static equilibrium thanks to the positive pressure of a cosmological constant. The geometry is perfectly symmetric under the exchange of the black holes and the flip of the sign of 
the charges.  However the scalar field is non constant and non symmetric, with different values at the horizons, which depend on a real modulus. Remarkably we show that it satisfies the attractor mechanism and the entropy indeed depends only on the magnetic 
charges. At one of the horizons the solution becomes $\frac 12$-BPS supersymmetric, while at the other one there is 
no supersymmetry, but the entropy remains independent from the scalar modulus.
\end{abstract}
\date{\today}
\maketitle

\section{Introduction}
Black holes are of great relevance in theoretical physics, and nowadays overwhelmingly entered in the realm of experimental physics, since the era of gravitational way spectroscopy started up. Their thermodynamical properties, like the 
microstate counting of the entropy, are the first trials by fire for any quantum theory. In supergravity theories, extremal black holes may represent supersymmetric solutions, \cite{Andrianopoli:2006ub} \cite{Mohaupt:2007mb} \cite{BPS2}, 
and, in gauged supergravities they can be seen as interpolating between a partially supersymmetric $AdS_2\times S^{d-2}$ topology at the horizon, the near horizon limit, and the fully supersymmetric anti-de Sitter solution at infinity
\cite{Cacciatori:2009iz} \cite{Klemm:2011xw} \cite{Chiara:2014ofa}. These are particularly important on the light of the conjectured duality between conformal quantum field theories on flat backgrounds and classical and quantum gravity in an 
anti-de Sitter space time, known under the name of AdS/CFT correspondence, \cite{AdS-CFT1} \cite{Zaffa1} \cite{Zaffa2}. It has numerous applications to QCD, condensed matter physics, hydrodynamics, superconductivity and in several cases goes 
beyond conformal field theories, \cite{Natsuume} \cite{Sachdev} \cite{Banks} \cite{Ling} \cite{Hartnoll} \cite{Iizuka} \cite{Hartnoll2} \cite{Aharony}.\\
In particular, black hole solutions in gauged supergravities are dual of condensed matter systems
at finite temperature \cite{Hartnoll}. In \cite{Iizuka} \cite{Charmousis} it is shown that black holes coupled with abelian gauge fields and neutral scalars can be used to describe transitions from fermi liquids to non fermi liquids, while when the scalar fields are 
charged the dual describes strongly coupled superconductors \cite{Hartnoll2}.
In recent years there have been progresses in finding BPS, non BPS and thermal black holes solutions in $N=2$ gauged supergravity in four dimensions, coupled with matter, see for example 
\cite{Hristov}, \cite{Gnecchi}, \cite{Cacciatori}
\cite{Daniele}, \cite{Klemm}, \cite{Chow-1}, \cite{Chow-2}, \cite{Chow-3}, \cite{Chong} and \cite{Hristov-1}. 
The work presented here belongs in this precise context, and is inspired by \cite{Astesiano:2020fwe}, where a new class of near horizon solutions, both stationary and static, in gauged supergravity coupled to scalar fields have been found, but they 
resisted to any effort in determining the black hole solutions of which they are the near horizon limit. In this work we fill this gap for the static case, and will show that a number of surprises arise. 
The stationary solution in \cite{Astesiano:2020fwe} belongs to a class of BPS solutions, which are everywhere regular with non constant scalar fields. More precisely, we considered the Fayet-Iliopoulos 
gauged supergravity coupled to three vector multiplets in the STU model with prepotential $F(X^0,X^1,X^2,X^3)=-2i\sqrt{X^0X^1X^2X^3}$ and the solutions were
characterized by an $AdS_2\times S^2$ geometry, and preserved four supersymmetry generators among the eight of the $N=2$, $d=4$ gauged supergravity coupled with four abelian gauge groups and three complex scalar fields. We will show that it is
indeed the near horizon limit of a black hole solution, which, however, is not supersymmetric at all: it consists in a pair of extremal black holes of equal sizes, facing symmetrically each other in a closed universe of positive curvature. It is a well known fact 
that a positive cosmological constant forbids supersymmetry. Multi black holes solutions in a de Sitter back ground are already known \cite{KT}; the Kastor-Traschen solution generalizes the Majumdar-Papapetrou multi black hole solution in asymptotically flat
background \cite{M} \cite{P}, but differ from the last, being non static. In our case, instead, the two black holes resemble more the black hole pairs of \cite{Mann} (and reference therein), but are in static equilibrium, despite, beyond gravitational attraction, 
they have to face also the electromagnetic attraction since they have opposite magnetic charges 
and zero electric charges, to form a sort of magnetic dipole. This attractive force is perfectly balanced by the pressure determined by a positive cosmological constant. The spatial section of such universe are represented by a finite space having the shape
of a three dimensional axially symmetric rugby ball at whose tips the curvature becomes singular. The two singularities are hidden by two horizon and the geometry is also symmetric under the reflection exchanging the tips. At the center of symmetry, 
the geometry is the one of a de Sitter space carrying a constant magnetic flux. However, while the same symmetry holds for the gauge fields (up to a sign), it does not extend to the scalar fields, which assume different values at the horizons. 
As a consequence, even though the geometry is the same, only at one of the horizon the full supersymmetric solution of \cite{Astesiano:2020fwe} is reproduced: the second horizon results to be non supersymmetric. It is worth to mention that the values 
of the scalar fields at the horizons depend on a real free modulus. On the other hand, the symmetric geometry is sufficient to ensure that both black holes have the same entropy, and the supersymmetry of one of them protect it from the dependence on 
the value of the scalar field at the horizon, t.i. on the scalar modulus. {Indeed, the solution turn out to satisfy the attractor mechanism at both horizons.}
The whole spacetime metric depends on three parameters, which can be interpreted as the positive cosmological constant, the value of the scalar modulus, and the magnetostatic
energy gap between the two horizons. 
All these properties make evident the peculiarity of this new solution.\\
We can now pass to the proof of our assertions.

\section{Equations of motion}
\noindent
We consider $N=2$, $D=4$ gauged supergravity coupled to $n$ abelian
vector multiplets.
Apart from the vierbein $e^a_{\mu}$, the bosonic field content includes the
vectors $A^I_{\mu}$ enumerated by $I=0,\ldots,n$, and the complex scalars
$z^{\alpha}$ where $\alpha=1,\ldots,n$. These scalars parametrize
a special K\"ahler manifold, i.~e.~, an $n$-dimensional
Hodge-K\"ahler manifold that is the base of a symplectic bundle, with the
covariantly holomorphic sections
\begin{equation}
{\mathcal V} = \left(\begin{array}{c} X^I \\ F_I\end{array}\right)\,, \qquad
{\mathcal D}_{\bar\alpha}{\mathcal V} = \partial_{\bar\alpha}{\mathcal V}-\frac 12
(\partial_{\bar\alpha}{\mathcal K}){\mathcal V}=0\,, \label{sympl-vec}
\end{equation}
where ${\mathcal K}$ is the K\"ahler potential and ${\mathcal D}$ denotes the
K\"ahler-covariant derivative. ${\mathcal V}$ obeys the symplectic constraint
\begin{equation}
\langle {\mathcal V},\bar{\mathcal V}\rangle = X^I\bar F_I-F_I\bar X^I=i\,. \label{sympconst}
\end{equation}
To solve this condition, one defines
\begin{equation}
{\mathcal V}=e^{{\mathcal K}(z,\bar z)/2}v(z)\,,
\end{equation}
where $v(z)$ is a holomorphic symplectic vector,
\begin{equation}
v(z) = \left(\begin{array}{c} Z^I(z) \\ \frac{\partial}{\partial Z^I}F(Z)
\end{array}\right)\,.
\end{equation}
F is a homogeneous function of degree two, called the prepotential,
whose existence is assumed to obtain the last expression.
The K\"ahler potential is then
\begin{equation}
e^{-{\mathcal K}(z,\bar z)} = -i\langle v,\bar v\rangle\,.
\end{equation}
The matrix ${\mathcal N}_{IJ}$ determining the coupling between the scalars
$z^{\alpha}$ and the vectors $A^I_{\mu}$ is defined by the relations
\begin{equation}\label{defN}
F_I = {\mathcal N}_{IJ}X^J\,, \qquad {\mathcal D}_{\bar\alpha}\bar F_I = {\mathcal N}_{IJ}
{\mathcal D}_{\bar\alpha}\bar X^J\,.
\end{equation}
The bosonic action reads
\begin{align}
e^{-1}{\mathcal L}_{\text{bos}} =& \frac 12R + \frac 14(\text{Im}\,
{\mathcal N})_{IJ}G^I_{\mu\nu}G^{J\mu\nu}  -\frac 18(\text{Re}\,{\mathcal N})_{IJ}\,e^{-1}
\epsilon^{\mu\nu\rho\sigma}G^I_{\mu\nu}G^J_{\rho\sigma} \nonumber
\\& -g_{\alpha\bar\beta}\partial_{\mu}z^{\alpha}\partial^{\mu}\bar z^{\bar\beta}
- V\,, \label{action}
\end{align}
with the scalar potential
\begin{align}
V = -2g^2\xi_I\xi_J[(\text{Im}\,{\mathcal N})^{-1|IJ}+8\bar X^IX^J]\,, \label{scal-pot}
\end{align}
that results from U$(1)$ Fayet-Iliopoulos gauging. Here, $g$ denotes the
gauge coupling and the $\xi_I$ are FI constants. In what follows, we define
$g_I\equiv g\xi_I$.
The Einstein's equations of motion from \eqref{action} are given by
\begin{align}
  \label{var-einstein}
  G_{\mu \nu} =&T_{\mu \nu}= ^{(0)}T_{\mu \nu} + ^{(1)}T_{\mu \nu}-g_{\mu \nu} V      \\
  ^{(0)}T_{\mu \nu} =& 2 g_{\alpha\bar \beta}\partial_{(\mu} z^\alpha \partial_{\nu)} \bar z^{\bar \beta}- g_{\mu \nu} g_{\alpha\bar \beta}\partial_\sigma z^\alpha \partial^\sigma \bar z^{\bar \beta} \label{stress-tensorspin0} \\
   ^{(1)}T_{\mu \nu} =& -(Im \mathcal N)_{IJ} G^I_{\mu\sigma} G_\nu^{J\sigma} + g_{\mu \nu}    \frac 14 (Im \mathcal N)_{IJ} G^I_{\sigma\rho} G^{J\sigma\rho} \label{stress-tensorspin1}
\end{align}
we make explicit the contribution form the spin $0$ and the spin $1$ parts. These equation give the curvature of the space-time
\begin{align}
   R = 2g_{\alpha\bar \beta}\partial_\sigma z^\alpha \partial^\sigma \bar z^{\bar \beta} +4 V,  \label{curvature1}
\end{align}
which can be used to rewrite the full system as
\begin{align}
R_{\mu\nu} &= -(\text{Im}\,{\mathcal N})_{IJ} G^I_{\mu\lambda} G^{J\ \lambda}_{\ \nu}  + 2 g_{\alpha\bar\beta} \partial_{(\mu}z^{\alpha}\partial_{\nu)}\bar z^{\bar\beta} + g_{\mu\nu}\left[ \frac{1}{4}  (\text{Im}\,{\mathcal N})_{IJ} G^I_{\rho\sigma} G^{J\rho\sigma}+ V\right] \,, 
\label{Einstein}
\end{align} 
\begin{align}
 \label{var-maxwell}
\nabla_\mu\left[(\text{Im}\,{\mathcal N})_{IJ} G^{J\mu\nu} - \frac{1}{2} (\text{Re}\,{\mathcal N})_{IJ} \, e^{-1} \epsilon^{\mu\nu\rho\sigma} G^J_{\rho\sigma}\right] = 0 ,\,   
\end{align}
\begin{align}
\label{var-scalars}
\begin{split}
\frac{1}{4} \frac{\delta(\text{Im}\,{\mathcal N})_{IJ}}{\delta z^\alpha} & G^I_{\mu\nu} G^{J\mu\nu} - \frac{1}{8} \frac{\delta(\text{Re}\,{\mathcal N})_{IJ}}{\delta z^\alpha} \, e^{-1} \epsilon^{\mu\nu\rho\sigma} G^I_{\mu\nu} G^J_{\rho\sigma} + \frac{\delta g_{\alpha\bar\beta}}{\delta \bar z^{\bar\gamma}} \partial_\lambda \bar z^{\bar\gamma} \partial^\lambda \bar z^{\bar\beta} \\
& + g_{\alpha\bar\beta} \nabla_\lambda \nabla^\lambda \bar z^{\bar\beta} - \frac{\delta V}{\delta z^\alpha} = 0 \,,
\end{split}    
\end{align}
which hold independently from the existence and the choice of a prepotential $F(X)$. \\
Defining the tensor
\begin{align}
    G_{I\mu \nu} = R_{IJ} G^I_{\mu \nu} + I_{IJ} \tilde{G}_{\mu \nu} \quad , \quad
    \tilde{G}^J_{\mu \nu} = \frac{1}{2} \sqrt{-g} \epsilon_{\mu \nu \rho \sigma} G^{J\rho \sigma},
\end{align}
then eq.(\ref{var-maxwell}), the Bianchi identities and the charges can be written as
\begin{align}
    \epsilon^{\mu \nu \rho \sigma} \partial_{\mu} \begin{pmatrix} G^I_{\rho \sigma}\\
  G_{I \rho \sigma}  \\ 
\end{pmatrix}=0  \quad , \quad
 \frac{1}{4\pi} \int_{\Sigma_{\infty}} \begin{pmatrix}  G^I\\
  G_I 
  \end{pmatrix}= 
  \begin{pmatrix}  p^I\\
  q_I\\
  \end{pmatrix}.
  \label{MaBi}
\end{align}
We now specialize to the ``magnetic'' $STU$ model, for which the prepotential can be written as
\begin{align}
 F=-2i\sqrt{(X^0X^1X^2X^3)}.
\end{align}
The symplectic section can be parametrised in terms of three complex scalar fields { $z^\alpha=\tau_\alpha$, $\alpha=1,2,3$,} so that {
\begin{align}
& v^T=e^{-\mathcal K /2}(X^J, \partial F/ \partial X^J)\cr &=(
 1, \tau_2 \tau_3, \tau_1 \tau_3, \tau_1 \tau_2, -i \tau_1 \tau_2 \tau_3, -i\tau_1,  -i \tau_2,  - i\tau_3 
).
\end{align} 
}
$\tau_\alpha$ and $\bar \tau_{\bar\alpha}$ are the complex coordinates on the scalar manifold.
The K\"ahler potential and the non vanishing components of the metric on the scalar manifold are respectively
\begin{align}
 e^{-\mathcal K} &=8 {\rm Re}\tau^1 {\rm Re}\tau^2 {\rm Re}\tau^3,  \\
 g_{\alpha\bar\alpha} &=g_{\bar\alpha\alpha}=\partial_{\alpha}\partial_{\bar\alpha}  \mathcal K=(\tau_{\alpha}+\bar\tau_{\bar\alpha})^{-2}.
\end{align}
{
The scalars $z^\alpha$ are coupled to the gauge fields via the period matrix $\mathcal N_{IJ}$ defined by
\begin{align}
 F_I&=\mathcal N_{IJ} X^J, \\
  \partial_{\bar \alpha} \frac {\bar F}{\bar X^I} -\frac 12 (\partial_{\bar \alpha} \mathcal K) \frac {\bar F}{\bar X^I} &=\mathcal N_{IJ} \left( \partial_{\bar \alpha} \bar X^J -\frac 12 (\partial_{\bar \alpha} \mathcal K)\bar X^J \right).
\end{align}}
In looking for a static solution, we now propose the ansatz
\begin{align}
   \tau_1 &= \sqrt{\frac{g_0g_1}{g_2g_3}} \tau(r), \quad \tau_2=\sqrt {\frac {g_0g_2}{g_1g_3}} k_2, \quad \tau_3=\sqrt {\frac {g_0g_3}{g_1g_2}} k_3,   \label{StaticTAnsatz} 
   \end{align}
 where $k_2$ and $k_3$ are real constants. The $X^I$ components of the symplectic section are
\begin{align}
    X^I  = \frac{1}{8} \sqrt{\frac{G}{8}} \left(\begin{array}{c} \frac{1}{|g_0|}\frac{1}{\sqrt{ {\rm Re} \tau k_2 k_3}} \\ \frac{1}{|g_1|}\sqrt{\frac{k_2 k_3}{ {\rm Re} \tau}} \\ \frac{1}{|g_2|}\sqrt{\frac{k_3}{ {\rm Re} \tau k_2}} \tau  \\ 
    \frac{1}{|g_3|}\sqrt{\frac{k_2}{ {\rm Re} \tau k_3}} \tau \end{array}\right)
\end{align}
where $G\equiv 64 \sqrt{g_0 g_1 g_2 g_3}$.
Since $\mathcal{N}_{IJ}= \frac{F}{2 (X^I)^2} \delta_{IJ}$, we obtain
\begin{align}
    \mathcal{N}_{IJ}=
 - i \frac{64}{G}
\begin{pmatrix}
g_0^2 k_2 k_3 \tau & 0 & 0 &0\\
0 & g_1^2 \frac{\tau}{k_2 k_3} & 0& 0 \\
0 & 0 & g_2^2 \frac{k_2}{ \tau k_3} &0 \\
0&0&0&g_3^2 \frac{k_3}{ \tau k_2} \\
\end{pmatrix} ,
\end{align}
 the choices $k_2 k_3>0$, together with Re$\tau>0$ imposed by K\"ahler geometry, guarantee
\begin{gather}
    {\rm Im}(\mathcal N_{IJ})<0.
\end{gather}
For the potential we have 
\begin{align}
    V     =& - \frac{G}{16} \left[ \frac{g_0}{|g_0|}\frac{g_1}{|g_1|} \frac{1}{{\rm Re}\tau} + \frac{g_0}{|g_0|}\frac{g_2}{|g_2|} \frac{1}{k_2}+ \frac{g_0}{|g_0|}\frac{g_3}{|g_3|} \frac{1}{k_3} \right.\cr
    &\left. + \frac{g_1}{|g_1|}\frac{g_2}{|g_2|} k_3+ \frac{g_1}{|g_1|}\frac{g_3}{|g_3|} k_2 + \frac{g_2}{|g_2|}\frac{g_3}{|g_3|} \frac{|\tau|^2}{{\rm Re}\tau} \right].
\end{align}
Now, from the equations of motion we get
\begin{align}
    R- 4 V = \frac{1}{2} g^{rr} \frac{(\partial_r \tau)( \partial_r \bar \tau)}{({\rm Re} \tau)^2}  . \label{curvature-1}
\end{align}
Usually, at infinity one expects the scalars to approach a constant value, so the r.h.s of eq.(\ref{curvature-1}) vanishes there, and spacetime should have constant curvature, $R_\infty=4V_\infty$, $V_\infty$ being determined by the asymptotic value of $\tau$.
For example, like in \cite{Klemm:2011xw}, or in \cite{Cacciatori:2009iz} (see (4.42), (4.43) and (4.44) therein), we can assume $g_I>0$. Assuming that at infinity the fields reach 
the extremum of the potential (at which $\tau=k_2=k_3=1$), we set $k_2=k_3=1$ and get
\begin{align}
    \tau_{ex}= 1, \qquad V_{ex}= -\frac{3G}{8}.
\end{align}
for the minimum of the potential $V$.
Thus, at infinity
\begin{align}
    R_{ex}=- \frac{3}{2} G<0,
\end{align}
so we expect an anti-de Sitter space. This is indeed what happens for the static version of the solutions in \cite{Klemm:2011xw} (see \cite{Cacciatori:2009iz}), which interpolate between a $AdS_2\times \Sigma$ $\frac 12$-supersymmetric geometry on the
horizon, and a full supersymmetric $AdS_4$. But things go very different for the new solutions in \cite{Astesiano:2020fwe}, where
$g_0, g_1<0$ and $g_2, g_3 >0$. In that case
\begin{align}
    V=\frac{G}{16 } \left[ -\frac{1 + |\tau|^2}{{\rm Re} \tau} +4 \right], \label{FI-potential}
\end{align}
so, the minimum of the potential and the corresponding curvature are
\begin{align}
    \tau&= 1, \qquad k_2= 1 , \qquad k_3=1, \\ 
    V_{ex}&= \frac{G}{8} \qquad R_{ex}= \frac{G}{2}, \label{extremal}
\end{align}
leading then to a de Sitter space! 
To look for the explicit solution having as near horizon limit the static solution in  \cite{Astesiano:2020fwe}, we make the ansatz
   \begin{gather}
   (ds)^2 = - A(r) dt^2 + A(r)^{-1} dr^2+  B(r) d\Omega_\kappa^2 ,\\
   G^I_{tr}= \frac{(\text{Im}\,{\mathcal{N}})^{-1|IJ}}{B(r)}  \left((\text{Re}\mathcal{N})_{JS} \, p^S - q_J \right) .
\end{gather}
where $d\Omega_{k}^2=d\theta^2+f_{\kappa}^2(\theta)d\phi^2$ is the metric on the two-dimensional surfaces $\Sigma=\{\mathrm{S}^2,\mathrm{H}^2\}$ of constant scalar curvature 
$R=2 \kappa $, with $\kappa=\{1,-1\}$, and
\begin{equation}
f_\kappa(\theta) = \frac{1}{\sqrt{\kappa}} \sin(\sqrt{\kappa}\theta) = 
\left\{\begin{array}{c@{\quad}l} \sin\theta\, & \kappa=1\,, \\                                             
                                             \sinh\theta\, & \kappa=-1\,. \end{array}\right.
\end{equation} 
The stress tensor for the spin-1 part $^{(1)}T_{\mu \nu }$ can be computed as 
\begin{align}
^{(1)}T^{t}_{\,t}=^{(1)}T^{r}_{\,r}=-^{(1)}T^{\theta}_{\,\theta}=-^{(1)}T^{\phi}_{\,\phi}= - \frac{1}{B^2}V_{BH}.
\label{tv1}
\end{align}
One can also check that
\begin{equation}
\frac{1}{4}\frac{\partial I _{IJ }}{\partial z^{a}}G_{\mu \nu
}^{I}G^{J |\,\mu \nu }- \frac{1}{8}\frac{\partial R _{IJ }}{\partial z^{a}}G_{\mu \nu }^{I }{^{\ast }{G%
}}^{J |\,\mu \nu }=-\frac{1}{B^2}\frac{\partial V_{BH}}{%
\partial z^{a}},  \label{scaltv1}
\end{equation}
where $\mathcal N_{IJ}\equiv R_{IJ}+i I_{IJ}$. We define the so-called Black Hole potential( \textit{BH potential})
\begin{equation}
V_{BH}\equiv -\frac{1}{2}(p^{I },\,q_{I })\left(
\begin{array}{cc}
I_{I J }+R_{I S }(I )^{-1|\,S T }R
_{T J } & \;-R _{I S }(I )^{-1|\,S J } \\
-(I )^{-1|\,I S }R_{S J} & \;(I )^{-1|\,I
J }
\end{array}
\right) \left(
\begin{array}{c}
p^{J } \\
q_{J }
\end{array}
\right).   \label{VBH-def}
\end{equation}
The field Maxwell and Bianchi field equations (\ref{MaBi}) are satisfied, while the Einstein's equations of motion (\ref{Einstein}) and the scalar field equation
\begin{align}
    R_{tt} &= \frac{A}{2B} \left( A'' B + A' B' \right) = \frac{A}{B^2} V_{BH} - A V \label{SRtt1}, \\
    R_{rr}&=- \frac{1}{2 A B^2} \left(A'' B^2 +A' B' B +2 A B'' B- A B'^2 \right)= - \frac{1}{A B^2} V_{BH} + \frac{1}{A} V + \frac{\tau' {\Bar{\tau}'}}{2(Re \tau)^2}, \label{SRrr1}\\
    R_{\theta \theta}&= -\frac{1}{2} \left(A'B' +A B''\right)+\kappa = \frac{1}{B}  V_{BH} +B V, \label{SRthth1}\\
    0&=\frac{1}{B^2} \partial_\tau V_{BH} + \partial_\tau V - \frac{1}{B} \frac{\left( B A \bar\tau'\right)'}{4(Re \tau)^2} + A \frac{({\bar \tau}')^2}{4 (Re \tau)^3}. \label{SS1}
\end{align}
Summing and subtracting eq.(\ref{SRtt1}) and (\ref{SRrr1}) we can rewrite the system as
\begin{align}
    \frac{ \left(2 B'' B-B'^2 \right)}{B^2} &= 4 \frac{(\sqrt{B})''}{\sqrt{B}}=- \frac{\tau' {\Bar{\tau}'}}{(Re \tau)^2},  \\
      A' B'+ A'' B= (A' B)'&= 2 \frac{V_{BH}}{B} - 2 B V, \\
      -\frac{1}{2} \left( A' B'+ A B'' \right) +\kappa= -\frac{1}{2} \left( A B' \right)' +\kappa&= \frac{V_{BH}}{B} + B V, \\
       0&=\frac{1}{B^2} \partial_\tau V_{BH} + \partial_\tau V - \frac{1}{B} \frac{\left( B A \bar\tau'\right)'}{4(Re \tau)^2} + A \frac{({\bar \tau}')^2}{4 (Re \tau)^3}. 
\end{align}
The system can be rewritten as
\begin{gather}
     \frac{ \left(2 B'' B-B'^2 \right)}{B^2} = - \frac{\tau' {\Bar{\tau}'}}{(Re \tau)^2}, \label{EQ1} \\
     A'' B - A B''= 4 \frac{V_{BH}}{B}-2\kappa,  \label{EQ2}\\
      (AB)''= -4 B V +2\kappa,  \label{EQ3}\\
      0=\frac{1}{B^2} \partial_\tau V_{BH} + \partial_\tau V - \frac{1}{B} \frac{\left( B A \bar\tau'\right)'}{4(Re \tau)^2} + A \frac{({\bar \tau}')^2}{4 (Re \tau)^3}. \label{EQ4}
\end{gather}
For the solution in \cite{Astesiano:2020fwe}, which we want to extend, we have 
\begin{align}
    p^I= \frac{1}{8g^I},\qquad q_I=0, \qquad  V_{BH}=\frac{1}{G} \left(\frac{1+|\tau|^2}{Re \tau}\right),\qquad
    \partial_\tau V_{BH}=\frac{1}{2G} \left(\frac{\Bar{\tau}^2-1}{(Re \tau)^2} \right), \qquad \kappa=1 \label{CA2S},
\end{align}
so we specify the equations to this case. Eq.(\ref{EQ4}) becomes
\begin{gather}
     0= \frac{1}{2G}\left[\bar \tau^2-1 \right] -B^2 \frac{G}{32} \left[ \bar \tau^2-1\right] - B \frac{\left( A B \bar\tau'\right)'}{4} + A B^2 \frac{({\bar \tau}')^2}{4 (Re \tau)}, 
\end{gather}
and using $\tau=f(r)+ i g(r)$, we get
\begin{gather}
    -\frac{1}{4} B \left(B A f'\right)'+\frac{A B^2
   \left(f'^2-g'^2\right)}{4
   f}+\left(\frac{1}{2 G}-\frac{1}{32} G
   B^2\right) \left(f^2-g^2-1\right)=0, \label{EQ4R}\\
   \frac{B}{4}  \left(B A' g'\right)'-\frac{A B^2
   f' g'}{2 f}-2  \left(\frac{1}{2
   G}-\frac{1}{32} G B^2\right)f g=0\label{EQ4I}.
\end{gather}
After imposing
\begin{gather}
    f(r)= \sqrt{1+g(r)(g_{H}-g(r))},
\end{gather}
where $g_{H}$ is the value at the horizon, eq.(\ref{EQ4R}) and (\ref{EQ4I}) becomes linearly dependent.

\section{The solution: Black hole dipole}
\noindent
Let us now present the full solution. The Fayet-Iliopoulos potential is given by (\ref{FI-potential}) and the electromagnetic fields and their duals are {
\begin{align}
    G^I=&\frac 1{8 B} \frac{{\rm Im} \tau}{ {\rm Re} \tau} \begin{pmatrix}
    -g_0^{-1}\\-g_1^{-1} \\ g_2^{-1} \\g_3^{-1}
    \end{pmatrix} dt \wedge dr + \frac{\sin{\theta}}{8} \begin{pmatrix}
    g_0^{-1}\\g_1^{-1}\\g_2^{-1} \\ g_3^{-1}
    \end{pmatrix}  d\theta \wedge d\phi,\\
    G_I &= - \frac{8 B^{-1}}{G {\rm Re} \tau} \begin{pmatrix}
    |\tau|^2 g_0\\|\tau|^2 g_1\\ g_2\\g_3
    \end{pmatrix} dt \wedge dr. \label{fieldnh}
\end{align}}
The Maxwell equations and Bianchi identities for the gauge fields are  already satisfied, while the Einstein's equations and the equations of motion for the scalars give the general solution
\begin{align}
    ds^2&= - \frac{G}{c^2} \frac{ (1-c^2 r^2)^2}{2 (1-c^2 r^2)+ G r_A^2} dt^2+\frac{4 c^2}{G} \frac{2 (1-c^2 r^2)+ G r_A^2}{(1-c^2 r^2)^2} dr^2\cr
    &\quad +4\frac {2 (1-c^2 r^2)+ G r_A^2}{G^2 r_A^2 } \left(d\theta^2 + \sin^2\theta \, d\phi^2\right),\label{FS} \\
   \tau_1 &= \sqrt{\frac{g_0 g_1}{g_2 g_3}} \frac{G r_A^2+2(1- c^2 r^2)-i2 \sqrt{2 G}  r_A c r}{2 (1+cr)^2+G r_A^2}, 
   \qquad \tau_2=\sqrt{\frac{g_0 g_2}{g_1 g_3}}, \qquad \tau_3=\sqrt{\frac{g_0 g_3}{g_1 g_3}},\\
G^I&=\frac{G^{5/2} r_A^3 }{\sqrt{2} \, 8} \frac{c r}{ \left(2(1-c^2r^2)+ G r_A^2 \right)^2} \begin{pmatrix}
    g_0^{-1}\\g_1^{-1} \\ -g_2^{-1} \\-g_3^{-1}
    \end{pmatrix}  dt \wedge dr + \frac{\sin{\theta}}{8} \begin{pmatrix}
    g_0^{-1}\\g_1^{-1} \\ g_2^{-1} \\g_3^{-1}
    \end{pmatrix}   d\theta \wedge d\phi,
\end{align} 
where $c$ is a constant assumed positive, and $r$ ranges in
\begin{align}
  r_{\leftarrow} \equiv  - \frac{1}{c} \sqrt{\frac{2+ G r_A^2}{2}}<r<\frac{1}{c} \sqrt{\frac{2+ G r_A^2}{2}} \equiv r_{\rightarrow}.
\end{align}
Notice that we rescaled the time coordinate $t\to t/2$ in order to get a more direct comparison with the near horizon limit in \cite{Astesiano:2020fwe}.
The curvature is everywhere regular with the exception of these two points, where it diverges at $+\infty$. The spatial section (at $\theta=\frac \pi2$) is therefore not embeddable in $\mathbb R^3$, but we can roughly think at it as a sort of
rugby ball with the singular points at the tips, where also the K\"ahler geometry becomes singular, since Re $\tau$ goes to zero there. 
The components $g_{tt}=-4 g^{rr}$ have two double roots at 
\begin{align}
 r_{\pm}=\pm \frac 1c, 
\end{align}
which represent two extremal event horizons.
\begin{figure}[h!]
\centering
\includegraphics[width=130mm]{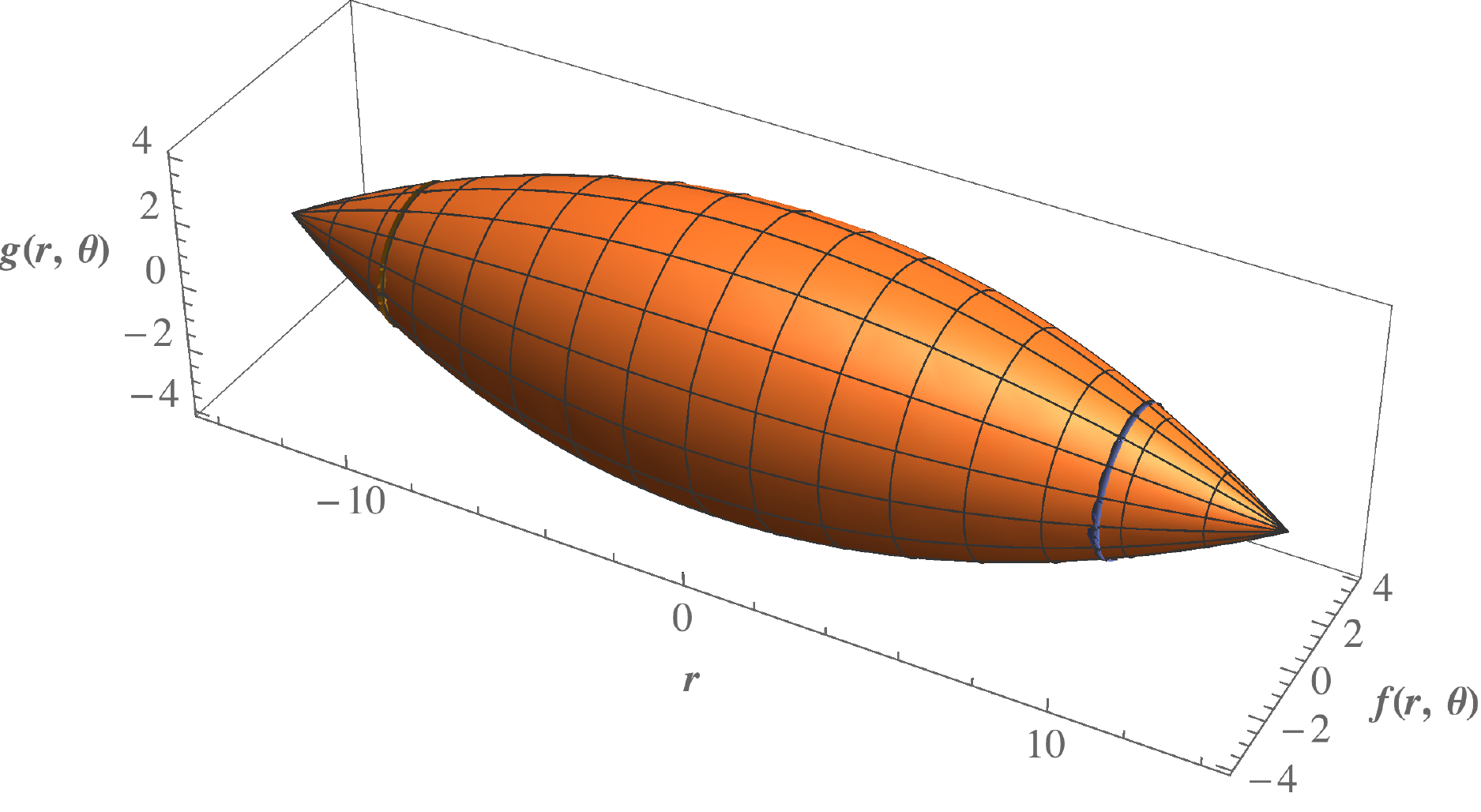}
\caption{Representation at $t,\phi$ constants of the geometry of the ``Black hole dipole"  with parameters $G=2,c=0.1,r_{A}=1$. The circles are the lines of constant $r$ and the longitudinal lines are the magnetic field lines. The horizons are at $r_{\pm }=\pm 10$ and are represented in blue, while the singularities are the tips at $r_{s\pm }=\pm 10 \sqrt{2}$. Here $f(r,\theta)=g_{\theta\theta} \cos{\theta} $ and $g(r,\theta)=g_{\theta\theta} \sin{\theta} $.}\label{Fig1}
\end{figure}
Both the horizons share the same geometry, since the limit at $r=\{r_-,r_+\}$ gives
\begin{align}
    ds^2=- 4\frac{r^2}{r^2_A} dt^2+ \frac{r_A^2}{r^2} dr^2+ \frac{4}{G}  d\Omega^2 , \label{NHL}
\end{align}
which is exactly the static near-horizon solution found in \cite{Astesiano:2020fwe}, after identifying 
\begin{align}
 r_A=\sqrt {\frac {2}{G}} \frac a\alpha.
\end{align}
In particular, the horizons have the topology of $AdS_2\times S^2$, with radii $|r_A|$ and $r_H=2/\sqrt G$ respectively.
Therefore, we have two black holes facing each other in a symmetric way, see Figure \ref{Fig1}. 
However, this symmetry under the parity transformation
\begin{align}
    r \rightarrow -r,
\end{align}
is not shared by the scalar $\tau_1$ and only up to a sign by the electromagnetic field, see Figure \ref{Fig2}. The fluxes have opposite signs, which means that the two black holes have opposite (magnetic) charges
\begin{align}
 p^I_{\mp}=\pm \frac 1{8g_I}, \qquad q^\mp_I=0.
\end{align}
Thus, they attract each other both gravitationally and electromagnetically.
\begin{figure}[h!]
\centering
\includegraphics[width=100mm]{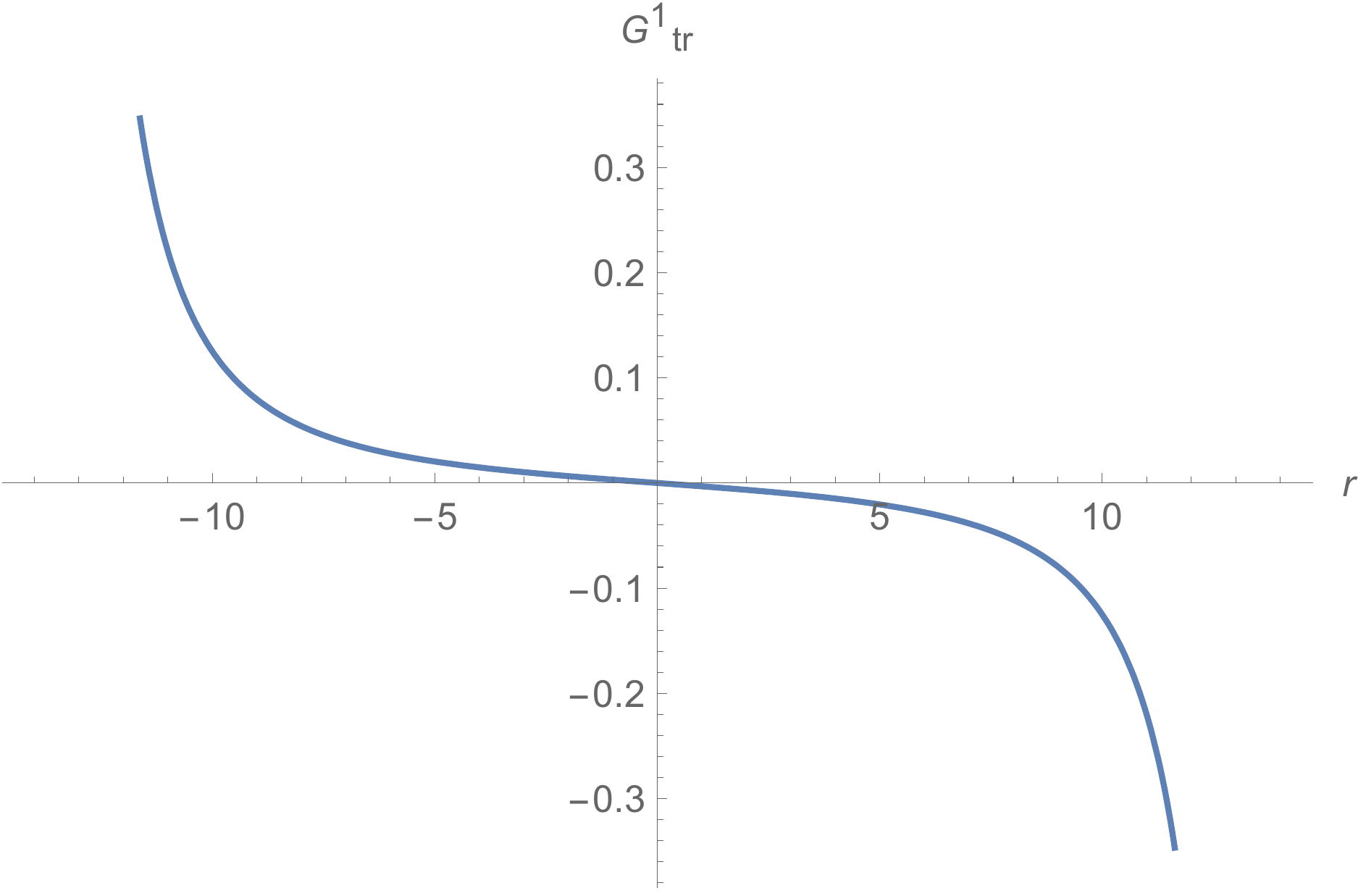}
\caption{Electric field interpreted as the $t,r$ component of $G^I$ for $G=2,c=0.1,r_{A}=1,g_1<0.$}  \label{Fig2}
\end{figure}
Why then they remain in static equilibrium? This can be understood noticing that the extremum (\ref{extremal}) of the potential is reached at $r=0$, and analyzing the metric around such region 
one checks that it approximate a de Sitter metric with cosmological constant $\Lambda=\frac G8$, and a magnetic flux $Q=\sqrt {\frac 2G}$, see Figure \ref{Fig3}.
\begin{figure}[h!]
\centering
\includegraphics[width=100mm]{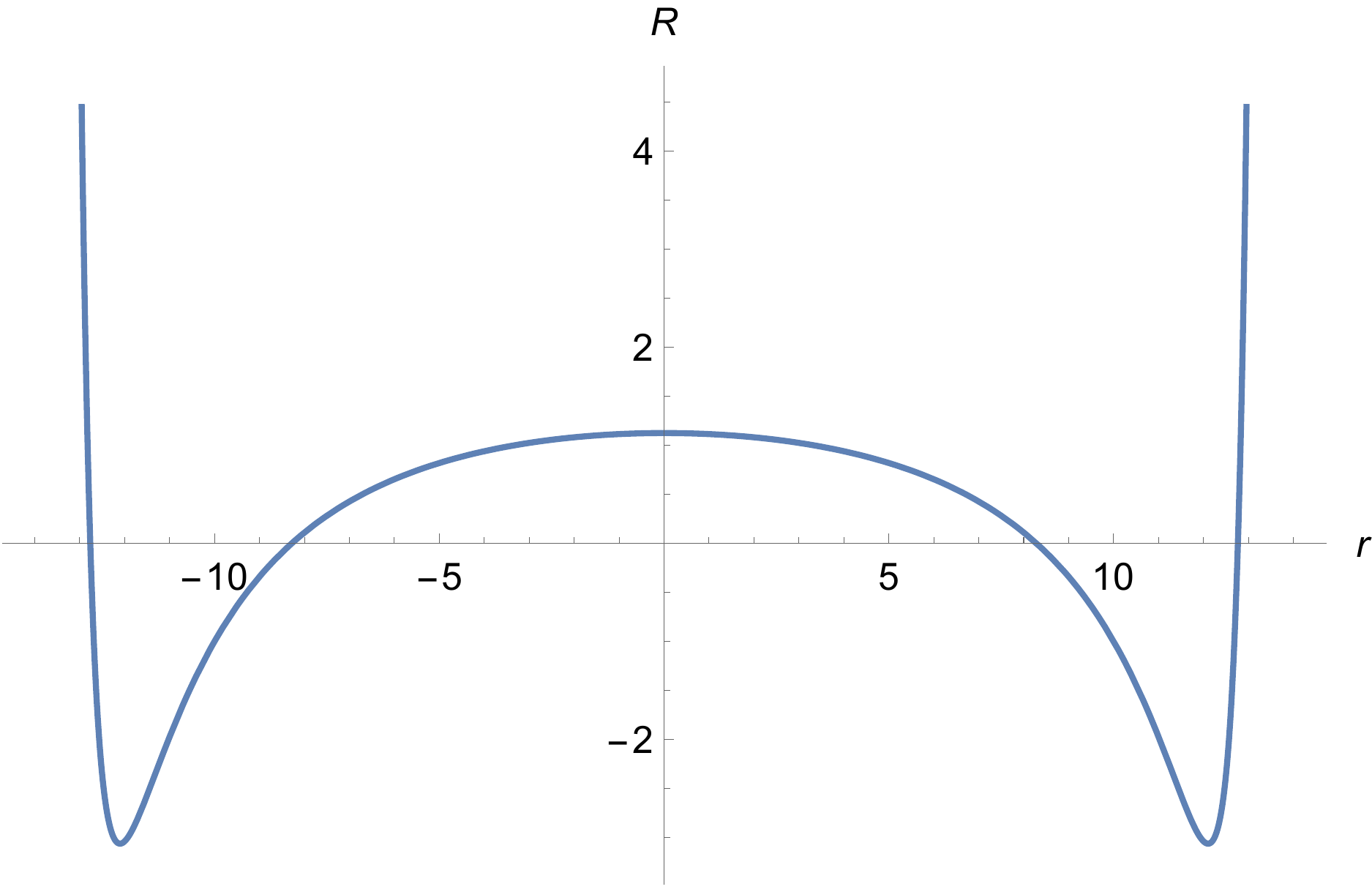}
\caption{Ricci scalar curvature for $G=2,c=0.1,r_{A}=1$.} \label{Fig3}
\end{figure}
It is interesting to compare the values of the non constant scalar at the horizons. One gets
\begin{align}
 \tau_1^+&=\frac {Gr_A^2}{8+Gr_A^2} \bar {\tau}_1^-, \quad
 \tau_1^-= \sqrt{\frac{g_0 g_1}{g_2 g_3}} \left (1+ 2i \frac{\alpha}a\right),
\end{align}
see Figure \ref{Fig4}. In particular, $\tau_1^-$ is exactly the scalar field of the supersymmetric solution in \cite{Astesiano:2020fwe} and we see that the horizon $r=r_-$ is $\frac 12$-BPS supersymmetric. Instead, the limit geometry at $r=r_+$ does not solve the
BPS equations and is not supersymmetric. 
\begin{figure}[h!]
\centering
\includegraphics[width=100mm]{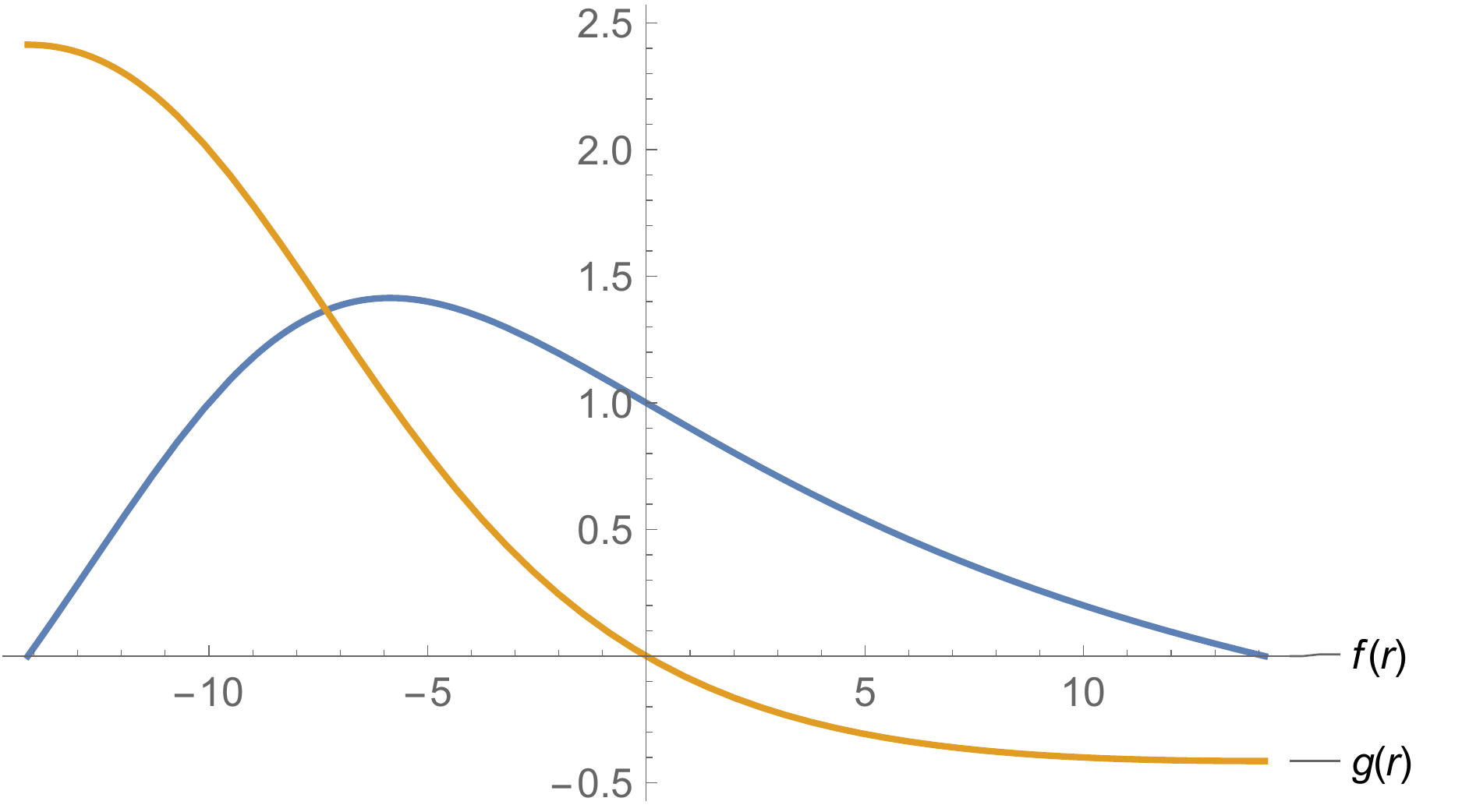}
\caption{Real (blue line) and imaginary part (orange line) of the scalar field with parameters $G=2,c=0.1,r_{A}=1.$} \label{Fig4}
\end{figure}
However, since the spacetime geometry is the same, the two black hole have the same entropy, which inherits some properties of supersymmetric black holes. Indeed,
\begin{align}
   S_{\pm}= \frac{4 \pi}{G}=4 \pi \sqrt{\prod_{I=1}^4 p^I_\pm},
\end{align}
which is independent on the values of the scalar at the horizons, which depend on the real modulus  $\alpha/a$. 
\\
The metric has four Killing vector fields, generating the four dimensional isometry group of rotations and time translations. This group is enhanced to $SO(2,1)\times SO(3)$ at the horizons. Since the solution is a closed universe, with no
asymptotic regions, the problem of correctly assigning a value to the masses of the black holes is a nontrivial task that needs a rather lengthy analysis, and that we demand to a separated work. However, it is worth to anticipate a couple of comments. 
First, from (\ref{fieldnh}) one easily gets the total electromagnetic energy separating the two horizons. For $G_I\equiv-d(\Upsilon_I\ dt)$ and $\Delta\Upsilon_I=\Upsilon_I(r_+)-\Upsilon_I(r_-)$, one gets 
\begin{align}
\Delta \mathcal E_\Upsilon=\frac 12\sum_I \Delta\Upsilon_I p^I=\frac 1c.
\end{align}
This gives a physical meaning to $c$, as a measure of the magnetostatic energy between the two black holes. Notice that it looks like a sort of (magnetic) capacitor whose walls are the horizons. The same value of the electromagnetic energy can be found
directly by integrating $-^{(1)}T_t^t$ on the spatial region between the two horizons. 
Second, despite we recognized a supersymmetric like characterization of the entropy, we expect a nontrivial contribution of the scalars to the first law of black holes thermodynamics \cite{GKK}. The computation of this contribution, 
however, is not trivial and requires an improved version of the method in \cite{Hajian} \cite{Hajian1}. \\
For extremal solutions the attractor mechanism implies for the scalars to be attracted toward the configuration in which, see \cite{Bellucci:2008cb},
\begin{gather}
\partial_\tau V_E|_{(r_H)}=0, 
\end{gather}
where $V_E$ is the effective potential,
\begin{gather}
    V_E= (1-\sqrt{1-4 V_{BH} V})/(2V).
\end{gather}
Remarkably, a direct check shows that in our case this is what happens at both horizons, and the entropy takes exactly the form
\begin{gather}
    S= \pi V_E(\tau_{rH})=  \frac{4\pi}{G}.
\end{gather}
It suggests that our solution behaves like a flow connecting a supersymmetric solution to a non supersymmetric one.
Making this flow more explicit, combined with AdS$_2$/CFT$_1$, would allow us to dualize a non-supersymmetric conformal field theory CFT$^{ns}_1$ to a supersymmetric conformal field theory CFT$^{s}_1$. 
\subsection{Comments on the mass and conserved quantities}
Since the dipole solution is closed and does not contain any asymptotically flat region, finding a good definition of mass for the contained black hoes is a quite hard problem.\\
A first idea to tackle this problem could be to notice that the region between the two horizons behaves as the throat of a wormhole. Using cut and sewing techniques one could get a true wormhole solution, as looked for in \cite{Maldacena:2020}. In our case we 
do not need to add a massless Dirac field, it is the stationary point of the potential that allows to sustain the wormhole. This interpretation can allow us to infer the mass of the two holes composing the wormhole. In that case one could be tempted to define the 
mass of each hole, as seen by an ``external'' observer, as the quantity measured by $M=\sqrt V_{BH}|_{Hor}$, the black hole potential at the horizon. Using the explicit values it would give
\begin{align}
 M=\frac {r_H}{r_A}\sqrt {\frac {r_H^2+r_A^2}2}. \label{MASS}
\end{align}
However, such a proposal is too naive and no further hint can help us in confirming such a result. Indeed, formula (\ref{MASS}) for the mass is usually true for solutions that admit an asymptotically flat or maximally symmetric solution, which is what 
would give rise a meaning to the word ``external'' referred to the observer. But the proposed mass formula does not result to satisfy any kind of Smarr relation, for example. Interestingly, the dipole solution has been found by us while looking for asymptotically 
flat or AdS solutions having the near horizon limit described in \cite{Astesiano:2020fwe}, but we failed in finding such a solution. So it is not at all evident that the above cut and glue operation can be really done in order to provide a meaning to our proposal.\\ 
One could try to define masses by looking at Komar integrals. Again, we have not an asymptotically flat region so that the usual results cannot be directly applied. Nevertheless, we could try with the following reasoning:\\
Let us review shortly the case of an asymptotically flat, for simplicity static, black hole, and let $K$ a timelike Killing vector field such that in the flat region, say $S_\infty$, $K^\mu K_\mu=-1$. Then, the mass of the black hole is
\begin{align}
 M=2\int_{S_\infty} \nabla^\mu K^\nu dS_{\mu\nu}.
\end{align}
On the other hand, using that $\nabla_\nu K^\nu=0$ and $[\nabla_\nu,\nabla^\mu]K^\nu=R^\mu_\nu K^\nu$, assuming that the black hole has an external event horizon at $S_+$, we can write
\begin{align}
 M=2\int_{S_+} \nabla^\mu K^\nu dS_{\mu\nu}+\int_\Sigma (2 T^t_\mu -\delta^t_\mu T) d\Sigma_t,
\end{align}
where $\Sigma$ is the spatial region external to the black hole. This formula is used in \cite{BaCaHa} to interpret the total mass of the black hole as the sum of the pure black hole mass
\begin{align}
M_{BH}= 2\int_{S_+} \nabla^\mu K^\nu dS_{\mu\nu}
\end{align}
plus the mass contribution from the matter outside the black hole
\begin{align}
 M_{matt}=\int_\Sigma (2 T^t_\mu -\delta^t_\mu T) d\Sigma_t.
\end{align}
For example, for a Reissner-Nordstrom black hole one would get\footnote{The $8\pi$ factor here arises from the convention $G_N=1/8\pi$ in place of $G_N=1$ as usual for the Newton constant.}
\begin{align}
 M_{BH}=8\pi\sqrt {M^2-Q^2}, \qquad M_{matt}=8\pi(M-\sqrt {M^2-Q^2}).
\end{align}
Notice that it is the total mass that enters the first law of thermodynamics. For an extremal black hole $M_{BH}=0$ and the total mass is completely given by the matter contribution. This may suggest us that, since our horizons are both extremal, the total mass
of each black hole could be given by the matter mass contribution of each half of the dipole external region (from $r_-$ to $r=0$ for the left horizon and from $r=0$ to $r_+$ for the right horizon). Indeed, a direct computation shows that for each of the horizons of
the dipole
\begin{align}
 \int_{S_\pm} \nabla^\mu K^\nu dS_{\mu\nu}=0.
\end{align}
Unfortunately, also 
\begin{align*}
 \int_{\{r=0\}} \nabla^\mu K^\nu dS_{\mu\nu}=0, \qquad \int_{r_-<r<0} (2 T^t_\mu -\delta^t_\mu T) d\Sigma_t=\int_{0<r<r_+} (2 T^t_\mu -\delta^t_\mu T) d\Sigma_t=0,
\end{align*}
so that this strategy is not able to give us any reasonable non vanishing mass for the black holes in the dipole.\\
The only interesting quantities we can associate to the dipole solution are the conserved ones associated to the Killing vectors describing the symmetries of the solutions. We have a timelike Killing vector $K_{(0)}=\partial_t$, associated to staticity, 
and the generators of the spherical symmetry. A direct computation show immediately that rotations give rise to vanishing integrals, so that the only non trivial conserved quantity is the total energy\footnote{Here we have included a factor $1/8\pi$ to easily
compare the results with the expressions given in standard units (for which the Newton constant is 1)}
\begin{align}
 E:=&-\frac 1{8\pi}\int_\Sigma T^t_t d\Sigma_t=-\frac 1{8\pi}\int_\Sigma {}^{(0)}T^t_t d\Sigma_t -\frac 1{8\pi}\int_\Sigma {}^{(1)}T^t_t d\Sigma_t+\frac 1{8\pi}\int_\Sigma V d\Sigma_t\cr
 =:& E^{(0)}+E^{(1)}+E_P.
\end{align}
Here $E^{(0)}$ is the kinetic energy of the scalar fields, $E^{(1)}$ represents the electromagnetic energy, and $E_P$ is the potential energy of the scalars. Explicitly:
\begin{align}
 E^{(0)}=& \frac{\left[4 \sqrt{G r_{A}^2+2}\left(7 G r_{A}^2+4\right)+ \sqrt{2} G r_{A}^2\left(3G r_{A}^2+8\right) \log \left(\frac{G r_{A}^2+4-2\sqrt{2G r_{A}^2+4}}{G r_{A}^2}\right) \right]}{4c G r_{A}^2 \sqrt{G r_{A}^2+2}} \\
 E^{(1)}=& \frac 1c, \\
 E_P=&E^{(1)}=\frac 1c.
\end{align}

\subsection{Geodesic motions}
Let us now study the geodesic equations in the black dipole background:
\begin{align}
    \ddot{r}&- \frac{G^2 r}{2 c^2} \frac{(1-c^2r^2)(1-c^2 r^2+G r_{A}^2)}{(2(1-c^2r^2)+Gr_{A}^2)^3} \dot{t}^2+ 2 c^2 r \left(\frac{1}{1-c^2r^2} -\frac{1}{2(1-c^2r^2)+Gr_{A}^2} \right) \dot{r}^2+\notag\\&+  \frac{2r}{Gr_{A}^2} \frac{(1-c^2r^2)^2}{2(1-c^2r^2)+G r_{A}^2} \left(\dot{\theta}^2+ \sin(\theta)^2 \dot{\phi}^2\right)=0, \label{r} \\
    \ddot{t}&+ 4 c^2 r \left(\frac{1}{1-c^2r^2} +\frac{1}{2(1-c^2r^2)+Gr_{A}^2} \right) \dot{t}\, \dot{r}=0,\label{t} \\
    \ddot{\theta}& - \frac{4c^2r}{2(1-c^2r^2)+Gr_A^2} \dot{\theta}\,\dot{r}- \cos(\theta)\sin(\theta) \dot{\phi}^2=0, \label{theta}\\
     \ddot{\phi}& - \frac{4c^2r}{2(1-c^2r^2)+Gr_A^2} \dot{\phi}\,\dot{r}+\frac{2}{\tan(\theta)} \dot{\theta}\dot{\phi}=0, \label{phi}
\end{align}
where $\dot{f}$ is the derivative respect to the affine parameter $\tau$. 
In the middle at $r=0$ there is a geodesics
\begin{gather}
    \theta,\phi=\text{constant}, t=\tau,
\end{gather}
which is an unstable equilibrium configuration. From the second and the fourth equations we directly get two constants of motion
\begin{gather}
    E= \dot{t} \frac{(1-c^2r^2)^2}{2(1-c^2r^2)+Gr_A^2},\label{E} \\
    L_z= \dot{\phi} \sin^2\theta \left[2(1-c^2r^2)+Gr_A^2\right], \label{M} 
\end{gather}
which can be interpreted as the energy (eventually per unit mass) and the $z$ component of the angular momentum.\\
After multiplying (\ref{theta}) times $2\dot \theta (2(1-c^2r^2)+Gr_{A}^2)^2$ and (\ref{phi}) times $2\dot \phi \sin^2\theta (2(1-c^2r^2)+Gr_{A}^2)^2$ and summing up the results we get
\begin{align}
 \frac {d}{dt} [(\dot\theta^2+\sin^2\theta \dot \phi^2) (2(1-c^2r^2)+Gr_{A}^2)^2]=0
\end{align}
so we get the constant of motion
\begin{align}
 L^2=(\dot\theta^2+\sin^2\theta \dot \phi^2) (2(1-c^2r^2)+Gr_{A}^2)^2, \label{L2}
\end{align}
representing the square modulus of the total angular momentum. Replacing from this last equation into (\ref{r}) we get
\begin{align}
    \ddot{r}&- \frac{G^2 E^2 r}{2c^2} \frac{1-c^2r^2+Gr_A^2}{(1-c^2r^2) \left[2(1-c^2r^2)+G r_A^2\right]}+\frac{2r}{G r_A^2}L^2 \frac{(1-c^2r^2)^2}{\left[2\left(1-c^2r^2\right)+G r_A^2\right]^3}+\notag\\ &  +2 c^2 r \dot{r}^2 \frac{1-c^2r^2+Gr_A^2}{(1-c^2r^2) \left[2(1-c^2r^2)+G r_A^2\right]}=0. \label{EOM}
\end{align}
This can be integrated giving a fourth integral of motion
\begin{align}
    - \frac{G}{c^2} \frac{(1-c^2r^2)^2}{2(1-c^2r^2)+Gr_{A}^2} \dot{t}^2+ \frac{4c^2}{G}\frac{2(1-c^2r^2)+Gr_{A}^2}{(1-c^2r^2)^2} \dot{r}^2 \cr+\frac{4}{G^2 r_{A}^2} \left[2(1-c^2r^2)
    +Gr_{A}^2\right] (\dot \theta^2+ \sin^2\theta \dot{\phi}^2)=-\epsilon, \label{IOM1}
\end{align}
where $\epsilon$ is 0 for massless particles and 1 in the massive case. This is of course the relation $g_{\mu\nu} \dot{x}^\mu\dot{x}^\nu=-\epsilon$. After substituting from the two integrals of motion ($\ref{E}$) and ($\ref{L2}$) we get
\begin{gather}
    -\frac{G}{c^2}E^2  \frac{2(1-c^2r^2)+Gr_{A}^2}{(1-c^2r^2)^2}+ \frac{4c^2}{G}\frac{2(1-c^2r^2)+Gr_{A}^2}{(1-c^2r^2)^2} \dot{r}^2+ \frac{4}{G^2r_A^2} \frac{L^2}{2(1-c^2r^2)+Gr_{A}^2}=-\epsilon.
\end{gather}
This equation can be equivalently rewritten as
\begin{gather}
    \frac{1}{2}\dot{r}^2+\tilde{V}(r)= \tilde{E}, \\
    \tilde{V}(r)= \frac{(1-c^2r^2)^2}{2(1-c^2r^2)+Gr_{A}^2}\left[\frac{L^2}{2c^2Gr_{A}^2}\frac{1}{2(1-c^2r^2)+Gr_{A}^2}+ \frac{G}{8c^2}\epsilon\right], \\
    \tilde{E}=\frac{G^2}{8c^4}E^2.
\end{gather}
\begin{figure}[h!] 
\centering
\includegraphics[width=100mm]{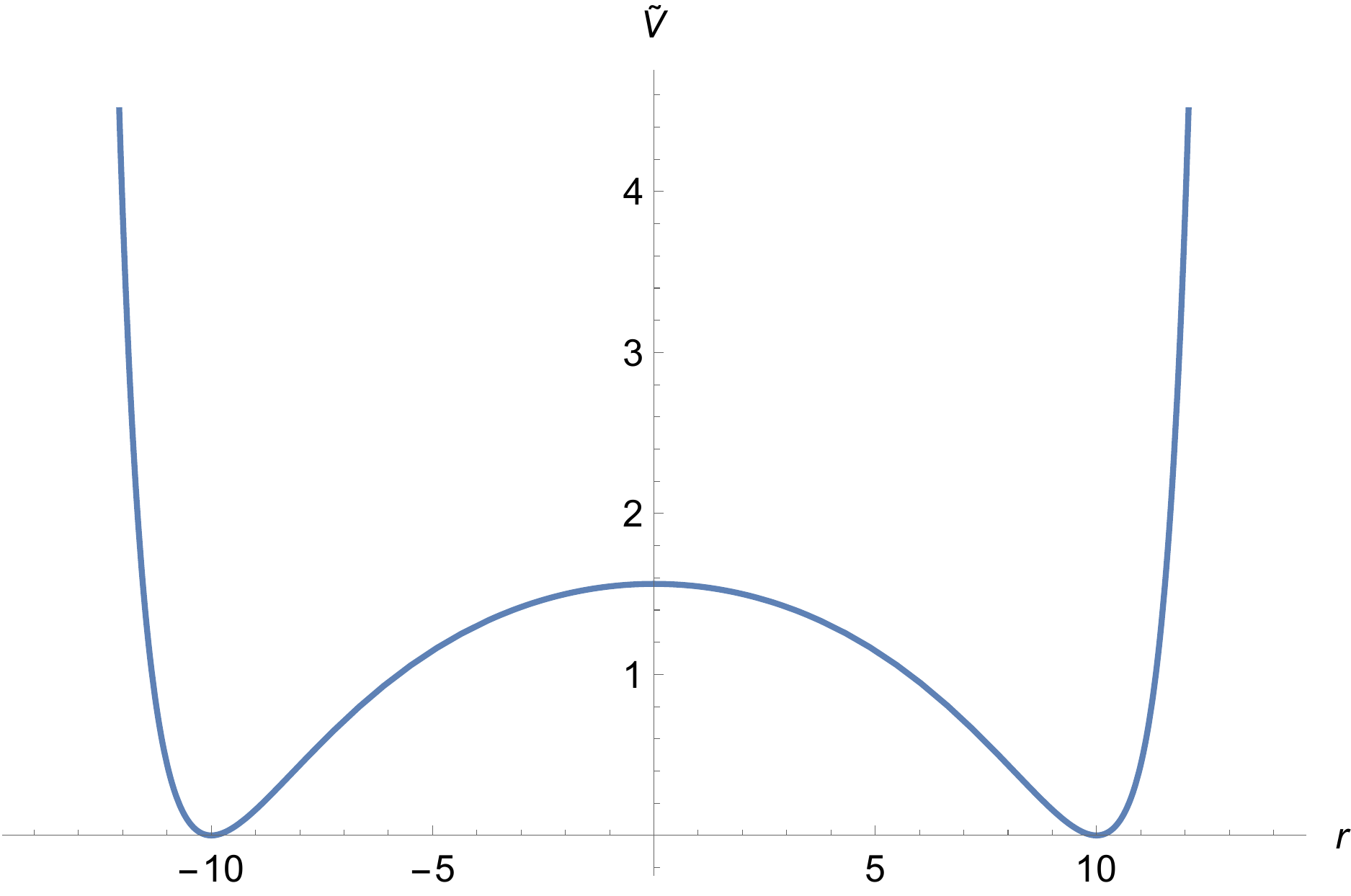}
\caption{$\tilde{V}$ with parameters $G=2,c=0.1,r_{A}=1,M=1,\epsilon=0.$}\label{POTFig}
\end{figure}
Thus, the problem of geodesic motions is completely reduced to quadratures.\\
For circular orbits we impose $r=r_c\neq 0$ in (\ref{EOM}) and obtain
\begin{gather}
    L^2= \frac{G^3 r_A^2 E^2}{4c^2} \frac{\left[2(1-c^2r_c^2)+Gr_{A}^2\right]^2\left(1-c^2r_c^2+Gr_{A}^2\right)}{(1-c^2r_c^2)^3},
\end{gather}
and if we then trade $E$ for $\epsilon$ using eq.(\ref{IOM1}) we get
\begin{gather}
    \epsilon=\frac{-4L^2}{G} \frac{1}{\left[2(1-c^2r_c^2)+Gr_{A}^2\right]\left(1-c^2r_c^2+Gr_{A}^2\right)},
\end{gather}
which is not compatible with $\epsilon \geq 0$, showing the non existence of circular orbits for $r\neq 0$, as one can expect from the form of $\tilde{V}(r)$ in fig.\ref{POTFig}. 
The conservation of the angular momentum implies that motions are planar. Indeed, let us set $\rho^2(r)\equiv \frac {4}{G^2r_A^2}2(1-c^2r_c^2)+Gr_{A}^2$ and let $u_r$, $u_\theta$, $u_\phi$ the spatial vierbein vectors so that the spatial position
of a test particle is given by the vector
\begin{align}
 \vec x=\rho u_r.
\end{align}
The spatial velocity, with respect to a parameter $\tau$, is $\vec v=\dot \rho u_r+\rho \dot u_r$. If we define the vector 
\begin{align}
 \vec \Lambda=\vec x \times \vec v,
\end{align}
we get immediately that
\begin{align}
 \vec \Lambda \cdot \vec \Lambda=\frac {16}{G^4r_A^4} L^2.
\end{align}
Moreover, let us define
\begin{align}
 u_z=\cos\theta u_r-\sin \theta u_\theta.
\end{align}
Then, it easily follows that
\begin{align}
 \Lambda_z:= \vec \Lambda \cdot u_z=\frac {4}{G^2 r_A^2} L_z.
\end{align}
We can then get an idea of the orbits by looking at the equations of motion on the plane $\theta=\pi/2$. In this case we easily get
\begin{gather}
    \phi(r)-\phi_0= \int \frac{dr}{\sqrt{\frac{2\tilde{E}}{L^2}\left[2(1-c^2r^2)+Gr_{A}^2\right]^2- \frac{(1-c^2r^2)^2}{c^2Gr_{A}^2} +\frac {G}{4c^2L^2}(1-c^2r^2)(2(1-c^2r^2)+Gr_{A}^2)\epsilon}},
\end{gather}
where $\epsilon=1$ for massive particles and $\epsilon=0$ for massless particles. This can be explicitly expressed in terms of elliptic function, but the explicit expressions would not give us major hints. \\
Notice that, as happens for Reissner-Nordstrom black holes, the singularities are repulsive. No geodesics fall in the singular region, but they are expelled from the future white hole.
\subsection{Causal structure}
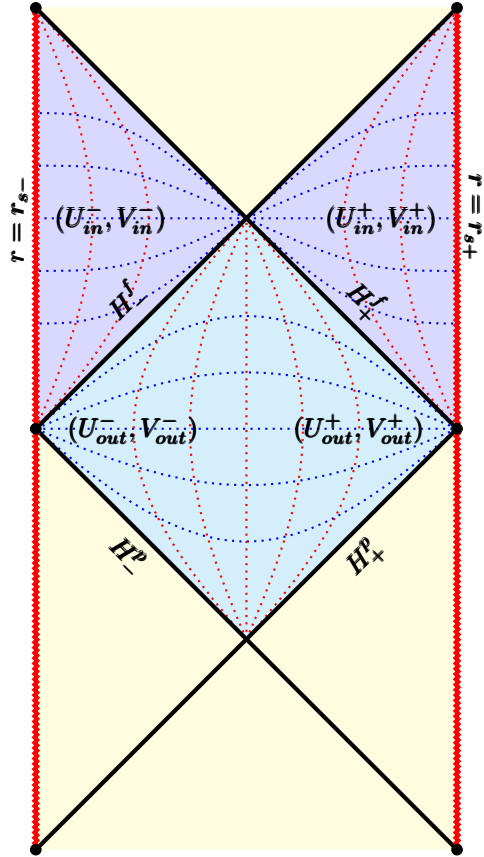
\begin{figure}[!h]
\begin{center}
\begin{tikzpicture}[>=latex,decoration={zigzag,amplitude=.5pt,segment length=2pt}]
\filldraw [yellow!15!white] (-2.8,-5.6) -- (-2.8,5.6) -- (2.8,5.6) -- (2.8,-5.6) -- cycle; 
\filldraw [cyan!15!white] (-2.8,0) -- (0,2.8) -- (2.8,0) -- (0,-2.8) -- cycle;
\filldraw [blue!15!white] (-2.8,0) -- (0,2.8) -- (-2.8,5.6) -- cycle;
\filldraw [blue!15!white] (2.8,0) -- (0,2.8) -- (2.8,5.6) -- cycle;
\draw [blue, dotted, thick] (-2.8,0) -- (2.8,0);
\draw [blue, dotted, thick] (-2.8,0) .. controls (-0.5,1) and (0.5,1) .. (2.8,0);
\draw [blue, dotted, thick] (-2.8,0) .. controls (-0.5,-1) and (0.5,-1) .. (2.8,0);
\draw [blue, dotted, thick] (-2.8,0) .. controls (-0.5,2) and (0.5,2) .. (2.8,0);
\draw [blue, dotted, thick] (-2.8,0) .. controls (-0.5,-2) and (0.5,-2) .. (2.8,0);
\draw [red, dotted, thick] (0,-2.8) -- (0,2.8);
\draw [red, dotted, thick] (0,-2.8) .. controls (1,-0.5) and (1,0.5) .. (0,2.8);
\draw [red, dotted, thick] (0,-2.8) .. controls (-1,-0.5) and (-1,0.5) .. (0,2.8);
\draw [red, dotted, thick] (0,-2.8) .. controls (2,-0.5) and (2,0.5) .. (0,2.8);
\draw [red, dotted, thick] (0,-2.8) .. controls (-2,-0.5) and (-2,0.5) .. (0,2.8);
\draw [blue, dotted, thick] (0,2.8) -- (2.8,2.8);
\draw [blue, dotted, thick] (0,2.8) .. controls (1.5,3.5) and (1.8,3.5) .. (2.8,3.5);
\draw [blue, dotted, thick] (0,2.8) .. controls (1.5,4.2) and (2.1,4.2) .. (2.8,4.2);
\draw [blue, dotted, thick] (0,2.8) .. controls (1.5,2.1) and (1.8,2.1) .. (2.8,2.1);
\draw [blue, dotted, thick] (0,2.8) .. controls (1.5,1.4) and (2.1,1.4) .. (2.8,1.4);
\draw [red, dotted, thick] (2.8,0) .. controls (1.8,2.3) and (1.8,3.3) .. (2.8,5.6);
\draw [red, dotted, thick] (2.8,0) .. controls (0.8,2.3) and (0.8,3.3) .. (2.8,5.6);
\draw [blue, dotted, thick] (0,2.8) -- (-2.8,2.8);
\draw [blue, dotted, thick] (0,2.8) .. controls (-1.5,3.5) and (-1.8,3.5) .. (-2.8,3.5);
\draw [blue, dotted, thick] (0,2.8) .. controls (-1.5,4.2) and (-2.1,4.2) .. (-2.8,4.2);
\draw [blue, dotted, thick] (0,2.8) .. controls (-1.5,2.1) and (-1.8,2.1) .. (-2.8,2.1);
\draw [blue, dotted, thick] (0,2.8) .. controls (-1.5,1.4) and (-2.1,1.4) .. (-2.8,1.4);
\draw [red, dotted, thick] (-2.8,0) .. controls (-1.8,2.3) and (-1.8,3.3) .. (-2.8,5.6);
\draw [red, dotted, thick] (-2.8,0) .. controls (-0.8,2.3) and (-0.8,3.3) .. (-2.8,5.6);
\draw [ultra thick] (-2.8,-5.6) -- (2.8,0) -- (-2.8,5.6);
\draw [ultra thick] (2.8,-5.6) -- (-2.8,0) -- (2.8,5.6);
\draw [ultra thick, red, decorate] (-2.8,-5.6) -- (-2.8,5.6);
\draw [ultra thick, red, decorate] (2.8,-5.6) -- (2.8,5.6);
\filldraw (-2.8,-5.6) circle (2pt);
\filldraw (2.8,-5.6) circle (2pt);
\filldraw (-2.8,5.6) circle (2pt);
\filldraw (2.8,5.6) circle (2pt);
\filldraw (-2.8,0) circle (2pt);
\filldraw (2.8,0) circle (2pt);
\node at (-1.5,0) {\pmb {$(U^{-}_{out},V^{-}_{out})$}};
\node at (1.5,0) {\pmb {$(U^{+}_{out},V^{+}_{out})$}};
\node at (-1.8,2.8) {\pmb {$(U^{-}_{in},V^{-}_{in})$}};
\node at (1.8,2.8) {\pmb {$(U^{+}_{in},V^{+}_{in})$}};
\node [rotate around={45:(-1.5,0.5)}] at (-1.5,0.5) {\pmb {$H^f_-$}};
\node [rotate around={-45:(1.5,0.5)}] at (1.5,0.5) {\pmb {$H^f_+$}}; 
\node [rotate around={-45:(-1.5,-0.5)}] at (-1.5,-0.5) {\pmb {$H^p_-$}};
\node [rotate around={45:(1.5,-0.5)}] at (1.5,-0.5) {\pmb {$H^p_+$}}; 
\node [rotate=90] at (-3,2.8) {\pmb {$r=r_{s-}$}}; 
\node [rotate=-90] at (3,2.8) {\pmb {$r=r_{s+}$}}; 
\end{tikzpicture}
\caption{The red dotted lines are constant $r$, the blue dotted lines are constant time. The central part is the exterior of the black holes. $H^f_{\pm}$ are the future horizons of the left ($-$) and right ($+$) black holes. 
$H^p_{\pm}$ are the past horizons. The darker sectors are the internal regions of the two black holes, the red waved lines are the timelike singularities. The yellow regions are periodic prolongations.}
\label{fig:Penrose}
\end{center}
\end{figure}

We want now to analyze the causal structure of the dipole black hole manifold.
Let us define new coordinates such that the $(t,r)$ part of the metric reads
\begin{gather}
    ds^2_{|t,r}= - \frac{G}{4c^2} \frac{ (1-c^2 r^2)^2}{2 (1-c^2 r^2)+ G r_A^2} \left(-dt^2+dr_{*}^2\right),
\end{gather}
where, compared to the metric $(\ref{FS})$, we defined a new time $t\rightarrow t/2$ and a new coordinate tortoise $r_{*}$
\begin{align}
    r_{*}= \frac{2 c^2 r_A^2 r}{1-c^2r^2}+ c\left(\frac{4}{G}+r_A^2\right) \log\left(\left|\frac{1+cr}{1-cr}\right|\right),
\end{align}
which takes values in $(-\infty,+\infty)$ as  $r \in (-\frac{1}{c},\frac{1}{c})$. While the metric $(\ref{FS})$ was well defined in the range $-\frac{1}{c}< r<\frac{1}{c}$, if we introduce the radial null coordinates as
\begin{gather}
    v= t+r_{*},\qquad u=t-r_{*}.
\end{gather}
the metric in the new coordinates $(v,r, \theta,\phi)$ 
\begin{gather}
ds^2= - \frac{G}{4c^2} \frac{ (1-c^2 r^2)^2}{2 (1-c^2 r^2)+ G r_A^2} dv^2+ 2 dv dr + \frac{4}{G^2 r_A^2}d\Omega^2,
\end{gather}
is smooth until the singularities $r_{s\pm}= \pm c \sqrt{\frac{2+Gr_A^2}{2}}$. To study the causal structure we introduce a first set of Kruskal coordinates
\begin{align} \label{KC1}
    -e^{- \frac{u}{2 c \left(\frac{4}{G}+r_A^2\right)}} \,:= & \begin{cases}
    U^{-}_{out} &\text{if}\quad r \in (-\frac{1}{c},\frac{1}{c}) \\
    U^{+}_{in}  &\text{if}\quad r \in (\frac{1}{c},r_{s+})
    \end{cases},
    \\  e^{ \frac{v}{2 c \left(\frac{4}{G}+r_A^2\right)}}\, := &\begin{cases}
    V^{-}_{out} &\text{if}\quad r \in (-\frac{1}{c},\frac{1}{c}) \\
    V^{+}_{in}  &\text{if}\quad r \in (\frac{1}{c},r_{s+}) \label{KC2}
    \end{cases}.
\end{align}
The products of these functions is
\begin{gather}
    UV= -e^{ \frac{c r_A^2}{\left(\frac{4}{G}+r_A^2\right)} \frac{r}{1-c^2 r^2}} \left(\frac{1+cr}{1-cr}\right),
\end{gather}
with positivity defined by 
\begin{gather}
     \frac{1+cr}{1-cr} >0 \rightarrow \quad r \in (-\frac{1}{c}, \frac{1}{c}).
\end{gather}
Even though the Kruskal coordinates are defined for $U<0$ and $V>0$ we can extend them to $U,V \in \mathbb{R}$ as usual, thinking $r=r(U,V)$. We have then 4 regions, two of them covered by the coordinates $(U^{-}_{out},V^{-}_{out})$ and $(U^{+}_{in},V^{+}_{in})$ respectively. The remaining two regions are covered by $(U^{-}_{in},V^{-}_{in})$ and $(U^{+}_{out},V^{+}_{out})$ defined as
\begin{align} 
    e^{ -\frac{v}{2 c \left(\frac{4}{G}+r_A^2\right)}} \,:= & \begin{cases}
    U^{-}_{in} &\text{if}\quad r \in (-\frac{1}{c},\frac{1}{c}) \\
    U^{+}_{out}  &\text{if}\quad r \in (\frac{1}{c},r_{s+})
    \end{cases},
    \\  -e^{ \frac{u}{2 c \left(\frac{4}{G}+r_A^2\right)}}\, := &\begin{cases}
    V^{-}_{in} &\text{if}\quad r \in (-\frac{1}{c},\frac{1}{c}) \\
    V^{+}_{out}  &\text{if}\quad r \in (\frac{1}{c},r_{s+}) 
    \end{cases}.
\end{align}
After compactifications the resulting Penrose diagram is illustrated in Figure \ref{fig:Penrose}.

\section{Conclusions}
\noindent
In \cite{Astesiano:2020fwe} we have found new supersymmetric backgrounds in $\mathcal N=2$, $D=4$ gauged supergravity coupled to vector multiplets for the STU model with prepotential $F(X^0,\ldots,X^3)=-2i\sqrt{X^0X^1X^2X^3}$. These are everywhere regular
solutions with one nonconstant complex scalar field and both magnetic and electric fields with vanishing fluxes. A parameter $\alpha$ can be set to zero to switch off the electric fields. 
In this limit the topology of the solutions is the one of a fibration $\mathbb R^{1,1}$ over a surface of genus $0$, while
for $\alpha\neq 0$ the solutions have the same topology as the Kerr-Newman throat: a fibration of $AdS_2$ fibres on the base of conformal spheres with non constant scalar curvature. These are stationary solutions, with a rotation parameter 
along a Killing spacelike direction on the conformal spheres. We recognized this solutions as near-horizon geometries of black holes. In this work we filled the gap in the static case and we found the alleged black hole. Surprisingly, the full solution presents a 
new interesting unusual black hole dipole. Among other properties, it is worth to mention that the scalars in this system satisfy the attractor mechanism. Indeed the entropy depends only on the charges.
In \cite{Astesiano:2020fwe} is considered also the case $\alpha=0$. In this case, the near horizon solution is a type II ultracold Nariai spacetime, which results to be the near horizon limit of a triply degenerate de Sitter spacetime
\begin{align}
 F^I=& \frac{\sin{\theta}}{8 g_I} d\theta \wedge d\phi,\quad    \tau =1, \\ 
    ds^2 =& - e^{2U(r)} dt^2 + e^{-2U(r)} dr^2 + r^2 \left(d\theta^2 +\sin^2{\theta} d\phi^2\right) \label{N}, \\ 
    e^{2U}=& -\frac{(\sqrt{G}r-2)^3(\sqrt{G}r+6)}{24 G r^2},
\end{align}
with horizon in $r=\frac{2}{\sqrt{G}}$, and magnetic charge $Q^2=\frac{2}{G}$, see e.g. \cite{Mann}.
It would be interesting to generalize these results to the rotating case and then analyze holography and the attractor mechanism along the lines of \cite{Anninos:2009yc} and \cite{Hotta}. First of all, in the rotating case we expect more non vanishing conserved
quantities that could allow us to find nontrivial relations helping us in determining the mass of the black holes. The second reason is that, as shown in \cite{Astesiano:2020fwe}, the shape of the horizon does not have constant curvature, so it looks like a deformed
sphere. This is a consequence of the presence of non trivial scalar configurations along the horizon. Therefore, in the rotating case we expect two possible scenarios: first, a dipole-like solution generalizing the above static solution, whose shape of the surfaces at constant $r$ is always the same when $r$ varies; second, an asymptotically AdS solution, whose shape of constant $r$ surfaces changes from a deformed sphere to a round sphere when moving from the horizon to the spatial infinity. This could help us in understanding if such deformations are allowed by the equations of motion and,
then, if open solutions, in place of the closed ones, exist or not according with the given family of near horizon solutions. In the case of affirmative answer, this should also provide a good definition of mass through the wormhole prolongation procedure we described in section 3.

\section*{Acknowledgments}
We thank an anonymous referee whose comments helped us to improve our manuscript.  
D.A. is thankful to Professor M. Trigiante and Professor A. Anabalon for useful discussions.




\newpage

\bibliographystyle{plain} 

\end{document}